\documentclass[12pt]{iopart}

\usepackage[dvips]{graphicx}
\usepackage{iopams}
\usepackage{cite}

\expandafter\let\csname equation*\endcsname\relax
\expandafter\let\csname endequation*\endcsname\relax
\usepackage{amsmath}
\usepackage{mathrsfs}

\renewcommand{\upsilon}{{\cal V}}

\newcommand{\ket}[1]{| #1 \rangle}
\newcommand{\bra}[1]{\langle #1 |}

\newcommand{\KK}{{G}}

\newcommand{\DD}{{\cal D}}
\newcommand{\ZZ}{{\cal Z}}
\newcommand{\Lg}{{\mathscr L}}
\newcommand{\Lgb}{\bar{\mathscr L}}

\begin{document}

\title{Coherent dynamics in long fluxonium qubits}

\author{Gianluca Rastelli,${}^{1,2}$ Mihajlo Vanevi{\' c}${}^3$ and Wolfgang Belzig${}^1$}
\address{${}^1$ Fachbereich Physik, Universit{\"a}t Konstanz, D-78457 Konstanz, Germany}
\address{${}^2$ Zukunftskolleg, Universit{\"a}t Konstanz, D-78457 Konstanz, Germany}
\address{${}^3$ Department of Physics, University of Belgrade, Studentski trg 12, Belgrade, Serbia}
\ead{gianluca.rastelli@uni-konstanz.de}

\begin{abstract}
We analyze the coherent dynamics of a fluxonium device [V.E. Manucharyan et al., Science
{\bf 326}, 113 (2009)] formed by a superconducting ring of Josephson
junctions in which strong
quantum phase fluctuations are localized exclusively on a single weak
element.
In such a system, quantum phase tunnelling by $2\pi$ occurring at the
weak element couples
the states of the ring with supercurrents circulating in opposite directions, while
the rest of the ring provides an intrinsic electromagnetic environment of the qubit.
Taking into account the capacitive coupling between nearest neighbors and the capacitance
to the ground, we show that the homogeneous part of the ring can sustain electrodynamic
modes which couple to the two levels of the flux qubit.
In particular, when the number of Josephson junctions is increased, several
low-energy modes can have frequencies lower than the qubit frequency.
This gives rise to a quasiperiodic dynamics which manifests itself as a decay of
oscillations between the two counterpropagating current states at short times, followed
by oscillation-like revivals at later times.
We analyze how the system approaches such a dynamics as the ring's length is increased
and discuss possible experimental implications of this non-adiabatic regime.
\end{abstract}


%
%
%
%
%
%
%
%
%
%
%
%
%
%
%
%

\date{\today}

\maketitle

\section{Introduction}
\label{sec:intro}

Quantum phase fluctuations in superconducting rings have attracted significant attention
during the last decade \cite{Mooij:1999,Orlando:1999,Friedman:2000,%
Caspar:2000,Wilhelm:2001,MLG:2002,Chiorescu:2003,Mooij:2005,Mooij:2006,%
Xue-Controllable-NJP07,Hausinger-Dissipative-NJP08,Manucharyan:2009,%
Manucharyan:2009-cond,Koch2009,Pop:2010,GuichardHekking:2010,%
Poletto-Coherent-NJP09,Hassler-Anyonic-NJP10,Vanevic:2011,Pop:2012,catelani_relaxation_2011,%
Manucharyan:2012,Astafiev:2012,peltonen_coherent_2013,Rastelli:2013,Susstrunk:2013,%
Spilla-Measurement-NJP14,Yamamoto-Superconducting-NJP14,Liu-Coexistence-NJP14}.
Embedding one or several Josephson junctions into a superconducting loop makes a
persistent current or flux qubit which enables the study of coherent quantum dynamics
between few quantum states, provided the system is sufficiently
decoupled from the external
environment \cite{Makhlin:2001,Devoret:2004,Devoret:2004-2,Clarke:2008}.
One of the first realizations of a flux qubit was achieved by using a superconducting loop
with a few Josephson junctions biased with an external magnetic field \cite{Mooij:1999}.
In such systems, two distinguishable macroscopic states with supercurrents circulating in
opposite directions exhibit oscillations due to quantum tunnelling.
Similarly, quantum phase fluctuations in superconducting
nanowires \cite{altomarePRL2006,Cirillo:2012,webster_nbsi_2013,hongisto_single-charge_2012}
can also exhibit coherent quantum dynamics when the wire is embedded
in a superconducting
loop threaded by an external magnetic
flux \cite{Mooij:2005,Mooij:2006,Vanevic:2011,Astafiev:2012,peltonen_coherent_2013}.
A possibility to realize analogous flux qubits in superfluid atom
circuits has been also analyzed recently \cite{Wright:2013,Amico:2013,Weiss:2015}.

%
%
%
%
\begin{figure}[b]
\begin{center}
\includegraphics[scale=0.54]{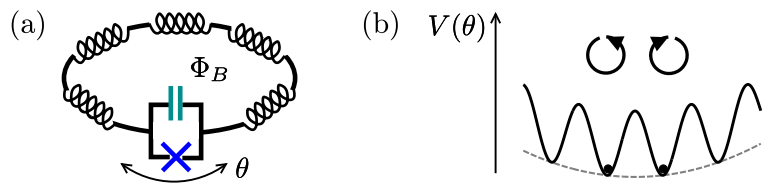}
\end{center}
\caption{(a) Schematic picture of a fluxonium qubit.
(b) Example of the potential for the phase difference across the single Josephson
junction in a pure inductive loop and for external magnetic flux $\Phi_B=\Phi_0/2$. The
two states associated with the local minima (black dots) correspond to the current
eigenstates of the loop with supercurrents circulating in opposite directions (arrowed
circles). Quantum phase tunnelling at the Josephson junction couples the two states.
}
\label{fig:fig1}
\end{figure}

For Josephson devices a particularly important achievement is the recent experimental
realization of the fluxonium qubit
\cite{Manucharyan:2009,Manucharyan:2009-cond,Koch2009,catelani_relaxation_2011,Manucharyan:2012}
in which coherent quantum phase tunnelling is localized at a single weak junction of the
Josephson chain.
As shown in figure~\ref{fig:fig1}, in such a device a small junction
is shunted with a series array
of $N \gg 1$ identical large-capacitance Josephson tunnel junctions.
The array of Josephson junctions acts as a superinductance which protects the small
junction from offset charge variations.
The junctions are characterized by a capacitance $C$, a maximal supercurrrent $I_J$,
the Josephson energy $E_J=\hbar I_J/2e$, and the charging energy $E_C=e^2/2C$.
Large phase fluctuations of order of $2\pi$ occur due to quantum
tunnelling with an amplitude
$\upsilon$ which in the limit $E_J \gg  E_C$ reads \cite{MLG:2002,catelani_relaxation_2011,Rastelli:2013,Likharev:1985}
\begin{equation}
\label{eq:upsilon}
\upsilon = 4 {(8 E_J^3 E_C/\pi^2)}^{1/4} \exp(-\sqrt{8E_J/E_C}).
\end{equation}
The fluxonium is realized for the condition
\begin{equation}
\label{eq:condition_1}
E_J/E_C \gg 1
\end{equation}
for $N$ junctions so that large phase fluctuations of order of $2\pi$ are
exponentially suppressed in the homogeneous part of the loop and each
Josephson junction
implements a linear inductance $L_J= \hbar^2/4e^2E_J$.
In contrast, the weak element is characterized by the parameters $\bar{E}_J$ and
$\bar{E}_C$ such that the amplitude of the phase tunnelling at the weak element
is $\bar{\upsilon}\gg \upsilon$.
More precisely, the relation
\begin{equation}
\label{eq:condition_2}
\bar{\upsilon} \gg N \upsilon
\end{equation}
holds in the fluxonium which ensures that the inductive role of the junction array is
not spoiled by large quantum phase fluctuations occurring in some part of it
\cite{Mooij:1999,Orlando:1999,MLG:2002,Manucharyan:2009,Rastelli:2013}.
We note that the condition \eqref{eq:condition_2} still allows for exponentially long
chains as long as  $\bar{E}_J/\bar{E}_C \ll E_J/E_C$.
Therefore, the non-linear excitations of the superconducting phase are
strongly localized in one part of the superconducting ring.
We remark that this is a special situation as, for instance, quantum phase slips in
homogeneous superconducting nanowires and vortex excitations require a non-perturbative
approach to treat the core of these non-linear excitations.

In this paper, we study coherent dynamics of the fluxonium device subject to an
externally applied magnetic flux $\Phi_B$ as a function of the size of the system.
We consider a one-dimensional ring composed of $N$ identical Josephson junctions
and a weak element where the strong phase fluctuations of order of $2\pi$ take place.
Such quantum phase fluctuations are localized {\it exclusively} at the weak element
whereas the rest of the chain acts as an electromagnetic environment, provided the
conditions \eqref{eq:condition_1} and \eqref{eq:condition_2} are satisfied.
We focus on the two-level regime for a magnetic flux close to a half flux quantum
$\Phi_B \approx \Phi_0/2$ ($\Phi_0=h/2e$) which is typical for
experimental flux qubits devices.
In this regime quantum tunnelling of the phase difference across the weak element
coherently couples the two states with supercurrents circulating in the opposite directions.
Taking into account the electrostatic interactions in the loop and in particular the
capacitance to the ground, the homogeneous part of the loop different from the weak
junction behaves as an ensemble of harmonic oscillators, similarly to electrodynamic
modes of a transmission line of a finite length.
We denote the spectrum of the modes by $\{\omega_k\}$, where $\omega_1$ is the
lowest frequency which scales with the size of the system as $\omega_1 \sim 1/N$.

We obtain the spectrum of electrodynamic modes $\{\omega_k\}$ in the ring and show that
the local phase difference $\theta$ across the weak element couples to these modes
which represent an effective (intrinsic) environment.
We find that the frequencies $\omega_k$ are not equidistant and the modes'
coupling to the phase $\theta$ is non-uniform, with low-energy modes being
more strongly coupled to $\theta$ than the high-energy ones.
There are two qualitatively different dynamic regimes depending on the size of the system.
For a small system, the frequency $\omega_1$ may be large such that the adiabatic
condition $2\bar\upsilon \ll \hbar \omega_1$ holds.
In this case, the dynamics is given by that of a two-level system of the qubit, that is,
it consists of quantum oscillations between the two counterpropagating supercurrent
eigenstates.
The effect of the high frequency modes is only the renormalization of the bare tunnelling
amplitude, $\bar{\upsilon} \to \tilde{\upsilon}$.
As the size of the system is increased, the frequencies of the modes
decrease. The resonant condition $2\tilde{\upsilon} = \hbar\omega_1$ is met for a certain
number of junctions $N^*$ in series, where $2\tilde{\upsilon}$ is the energy splitting
between the first excited state and the ground state of the qubit.
For $N>N^*$ the system enters the non-adiabatic regime in which some modes have
frequencies smaller than the level splitting, $\hbar \omega_k < 2\tilde{\upsilon}$ for
$k=1,\dots,n$.
In this case, we find that the quantum dynamics is not periodic: it exhibits decay of
oscillations at short times, followed by revival-like oscillations at longer times.

The remainder of the paper is organized as follows.
In section~\ref{sec:QPS}, we recapitulate some of the results for the coherent phase
tunnelling in the fluxonium qubit.
Then, in order to facilitate a physical understanding of the effect of intrinsic electric
modes on the quantum phase dynamics of the single junction, in section~\ref{sec:model} we
use a semi-analytical approach for a generic model of a particle in a double well
potential coupled to a finite discrete bath of harmonic oscillators.
We find that the dynamics of the particle has qualitatively different regimes (coherent
and quasiperiodic) depending on the ratio between the tunnelling amplitude and the
frequencies of the oscillators ($\hbar\omega_1 \gg 2\tilde{\upsilon}$ and  $\hbar\omega_1
\lesssim 2\tilde{\upsilon}$).
In section~\ref{sec:1DJJ}, we take into account the electrodynamics of the loop and map
the fluxonium device to the model of section~\ref{sec:model}.
We obtain the frequencies and the coupling strengths of electric modes of the ring and we
analyze the experimental feasibility to observe the non-adiabatic dynamics.
In section~\ref{sec:Summary} we present our conclusions.

\section{The model for the fluxonium}
\label{sec:QPS}
%
%
%
%
%
%
\begin{figure}[t]
\begin{center}
\includegraphics[scale=0.23]{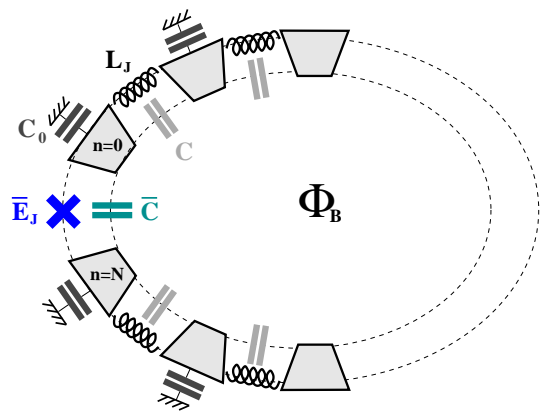}
\end{center}
\caption{(colors online) Superconducting ring made of $N$ identical tunnelling junctions
with inductance $L_J$ and capacitances $C$, and a weaker Josephson junction with
Josephson energy  $\bar{E}_J$ and capacitance $\bar{C}$. The ground capacitance of the
superconducting islands between the junctions is $C_0$.}
\label{fig:JJring}
\end{figure}

\subsection{Single Josephson junction in an inductive loop}

Let us consider a superconducting ring which consists of $N$ identical Josephson junctions
and a weaker Josephson junction at which the strong phase fluctuations are localized, as
has been discussed in the introduction.
The system is shown in figure~\ref{fig:JJring}.
If one neglects the electrostatic interactions in the homogeneous part of the loop,
the array formed by $N$ identical Josephson junctions provides an inductance
$L  = N L_J$.
This inductance sets the inductive energy $E_L = \hbar^2/4e^2 L$.
In this regime, the Hamiltonian of the system reads
\cite{Manucharyan:2009,Koch2009,catelani_relaxation_2011}
\begin{equation}
\label{eq:fluxonium_adiabatic}
\hat{H} = 4 \bar{E}_C \hat{n}^2 - \bar{E}_J \cos(\hat{\theta})
+\frac{E_L}{2} {\left( \hat{\theta} - 2\pi \frac{\Phi_B}{\Phi_0}
\right)}^2,
\end{equation}
where $\hat{\theta}$ is the operator of the phase difference across the weak junction
and $\hat{n}$ is its canonically conjugate operator ($[\hat{\theta},\hat{n}]=i$) that
represents the number of Cooper pairs that have passed across the junction.
Quasiparticle excitations and their dynamics can be disregarded
at very low temperatures $k_BT \ll \Delta$, where $\Delta$ is the
superconducting gap of the islands forming the Josephson ring.
In general, non-equilibrium distribution and trapped quasiparticles can give rise to the
decoherence of the flux qubit \cite{catelani_relaxation_2011}.

The spectrum of the Hamiltonian \eqref{eq:fluxonium_adiabatic} was analyzed
in \cite{Koch2009}.
The second and the third term in \eqref{eq:fluxonium_adiabatic} represent the energy
potential for the phase difference $\theta$.
An example is shown in figure~\ref{fig:fig1}(b) for $\Phi_B=\Phi_0/2$ where the two
absolute minima of the potential are degenerate and the two states are the
eigenstates of the supercurrent circulating in the opposite directions in the loop.
These states are distinguishable as they have different supercurrents.
For large inductance $E_L \ll \bar E_J$, the two classical minima of the phase are
given approximatively by $\pi E_L/\bar E_J$ and $2\pi(1-E_L/2\bar E_J)$ with the associated
supercurrents $\pm \Phi_0/4\pi L$.
The first term in \eqref{eq:fluxonium_adiabatic} is the electrostatic energy of the
weak junction and it plays the role of inertial kinetic energy such that quantum tunnelling
of the local phase difference $\theta$ can occur from one minimum to the nearest
neighbor wells in the phase potential as shown in figure~\ref{fig:fig1}.
The tunnelling amplitude can be obtained using semiclassical instanton or
Wentzel-Kramers-Brillouin (WKB) method in the regime $\bar{E}_C \ll \bar{E}_J$.
For $E_L \ll \bar E_J$, the profile of the energy barrier separating the two classical
states is well approximated by the cosine potential and the tunnelling amplitude
$\bar{\upsilon}$ is still given by equation \eqref{eq:upsilon} with $E_J$ and $E_C$
replaced by $\bar{E}_J$ and $\bar{E}_C$ of the weak element.
Corrections due to the parabolic part of the potential stemming from the loop inductance
were discussed in \cite{Rastelli:2013}.

For $E_L \gg 2 \bar{\upsilon}$, the quantum tunnelling effectively couples the two
neighboring minima corresponding to the current eigenstates and the low-energy
effective Hamiltonian can be described by a two-level model,
\begin{equation}
\label{eq:H_two_levels}
\hat{H} =
E_1 \left| 1 \right> \left< 1 \right|
+
E_2 \left| 2 \right> \left< 2 \right|
 -
 \bar{\upsilon}\left( \left| 1 \right> \left< 2 \right| +  \left|  2 \right> \left< 1  \right|  \right),
\end{equation}
where $E_{i} = (E_L/2) {(\theta_i - 2\pi \Phi_B/\Phi_0 )}^2$ and $\theta_i$ are the positions
of the two minima.
Although we implicitly  assumed $\Phi_B=\Phi_0/2$ for the derivation of the effective
Hamiltonian \eqref{eq:H_two_levels}, the obtained result is also applicable in a small
range of fluxes centered around the half flux quantum for which the energy difference
remains small as compared to the tunnelling amplitude, $|E_1-E_2| \ll \bar{\upsilon}$.
Thus, quantum tunnelling of the Josephson junction phase difference couples the
counterpropagating supercurrent states, giving rise to the
avoided crossing of the energy levels.
%
The level splitting at the degeneracy point is $E_S=2 \bar{\upsilon}$.
The splitting has been observed in fluxonium junctions close to degeneracy point
\cite{Manucharyan:2009,Manucharyan:2012} and in superconducting
nanowires \cite{Astafiev:2012,peltonen_coherent_2013}.

\subsection{Single Josephson junction coupled to the electric modes of the loop}

In this section we consider the superconducting Josephson ring taking into account the
electrostatic interaction in the homogeneous part as shown in figure~\ref{fig:JJring}.
Let $\phi_n$ be the superconducting phases of $N+1$ islands forming the ring.
We denote the phase difference across the weak junction by $\theta=\phi_0-\phi_N$
whereas $\theta_n = \phi_{n+1}-\phi_n$ ($n=0,\dots,N-1$) are the phase differences
across the Josephson junctions in the ring.
The corresponding Euclidean Lagrangian (in the imaginary time $\tau$) of the system
reads \cite{Rastelli:2013,Susstrunk:2013,Korshunov:1986,Korshunov:1989,Masluk:2012}
%
%
%
%
%
%
%
\begin{equation}
\label{eq:1DJJ_1}
\Lg  =
\sum_{n=0}^{N-1}
\left[  \frac{\hbar^2\dot\theta_n^2(\tau)}{16E_C}
+
\frac{\Phi_0^2}{2L_J}    {\left( \frac{\theta_n(\tau)}{2\pi} + \frac{\Phi_B}{(N+1)\Phi_0}   \right)}^2 \right]
+
\sum_{n=0}^N \frac{\hbar^2 \dot\phi_n^2(\tau)}{16E_0} +  \Lgb,
\end{equation}
%
%
%
%
%
%
where the Euclidean Lagrangian of the weak element is
\begin{equation}
\label{eq:1DJJ_2}
\Lgb = \frac{\hbar^2 \dot\theta^2(\tau)}{16\bar{E}_C}
- \bar E_J \cos\left( \theta(\tau) + \frac{2\pi\Phi_B}{(N+1)\Phi_0} \right).
\end{equation}
Here, $\dot\theta =d\theta/d\tau$,
$E_C=e^2/2C$ and
$E_0=e^2/2C_0$ are the charging energies of the islands
with
$C$ being the junction capacitance and $C_0$ the capacitance
of the islands to the ground.
The weak junction is characterized by $\bar{E}_J,\bar{E}_C$
where $\bar{E}_J < E_J$ and  $\bar{E}_C > E_C$.
The phases $\theta_n +  2\pi(\Phi_B/\Phi_0)/(N+1)$ in \eqref{eq:1DJJ_1} and
\eqref{eq:1DJJ_2} are the gauge invariant phase differences across the junctions.
Due to the phase periodicity $\phi_n = \phi_{n+N+1} + 2\pi m$ ($m$ integer), the variable
$\theta$ and the set of $N$ phase differences $\{ \theta_n \}$ satisfy the
constraint
\begin{equation}
\label{eq:constraint}
\theta(\tau) + \sum_{n=0}^{N-1}  \theta_n(\tau) =  0 \quad \left( \rm{mod\ } 2\pi \right).
\end{equation}
Using the path-integral formalism, one can write the partition function of the system as
\begin{equation}
\ZZ = \oint \DD\theta \prod_{n=0}^{N-1} \oint \DD\theta_n \exp\left(- \frac{1}{\hbar} \int^{\beta}_0 d\tau \Lg
\right)
\end{equation}
where $\beta=\hbar/(k_BT)$.

Before concluding this section, we note that equation \eqref{eq:fluxonium_adiabatic} can
be simply recovered from \eqref{eq:1DJJ_1} and \eqref{eq:1DJJ_2} by neglecting the
electrostatic interaction for the $N$ junctions in the chain and using the constraint
\eqref{eq:constraint} for the phase difference to impose $\theta_n= \theta / N $
in the limit $N\gg 1$.


\section{Effective model}
\label{sec:model}

Before we proceed with the study of dynamics of the quantum phase tunnelling across the
weak element  coupled to electric modes of the loop, in this section we first
analyze a generic model of a particle in  a double-well potential interacting with a
discrete bosonic bath. Mapping of the Josephson junction chain to this model is given in
section~\ref{sec:1DJJ}.

\begin{figure}[t]
\begin{center}
\includegraphics[scale=0.2,angle=0.]{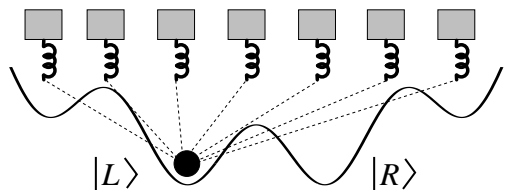}
\end{center}
\caption{Particle in a double-well potential coupled to $N$
harmonic oscillators. States localized around the two potential
minima are denoted by $\ket{L}$ and $\ket{R}$, respectively.}
\label{FIG_1dw_oscillators}
\end{figure}

Let us consider a particle moving in a double-well potential and
interacting with a bosonic bath of $N$ harmonic oscillators, see
figure~\ref{FIG_1dw_oscillators}.
If the height of the barrier is larger than the kinetic energy
$E\sim\hbar^2/ma_0^2$,
the system can be reduced to the states $\ket{R}$ and $\ket{L}$
localized at the positions $x=\pm a_0$
which are coupled by quantum tunnelling. This is a well-known
spin-boson model\cite{Leggett:1987,Weiss:2012,Breuer:2007}
with the Hamiltonian
\begin{equation}
\label{eq:spin-boson}
\hat H
=
\bar\upsilon \, \hat\sigma_x
+
\hat\sigma_z
\sum_{k=1}^N \alpha_k \hbar \omega_k
\left(  \hat a_k^\dagger + \hat a_k \right)
+
\sum_{k=1}^N \hbar \omega_k \hat a_k^\dagger \hat a_k.
\end{equation}
Here, $\hat\sigma_x = \ket{L}\bra{R} + \ket{R}\bra{L}$,
$\hat\sigma_z = \ket{L}\bra{L} - \ket{R}\bra{R}$,
$\hat a_k^\dagger$ ($\hat a_k$) are creation (annihilation)
operators of the oscillator modes $\omega_k$, $\alpha_k$ are
the coupling constants, and $\bar\upsilon$ is the bare
tunnelling amplitude between the states $\ket{L}$ and $\ket{R}$.
%
%
%
%
%
%
The two energy-degenerate states correspond to the
counterpropagating supercurrent states at half flux quantum as
discussed in section~\ref{sec:QPS}, whereas the harmonic oscillators
represent electric modes of the homogeneous part of the
superconducting loop in which large phase fluctuations are suppressed as discussed in
section~\ref{sec:intro}.
The coupling constants are related to the characteristic impedance
of the homogeneous part of the loop, see section~\ref{sec:1DJJ}.

For a large number of oscillators and linear low-frequency dispersion
($N\to\infty$, $\delta\omega\to 0$, $\omega_k = k\delta\omega$)
one recovers the standard Caldeira-Leggett model\cite{Caldeira:1981}
which describes the dissipative quantum dynamics of the two-level
system coupled to an ohmic environment.
This system has been studied extensively in the
literature \cite{Caldeira:1981,Leggett:1987,Weiss:2012,Breuer:2007}.
Here we just recall that
%
the high-energy modes with $\hbar\omega_l \gg 2\bar\upsilon$
quickly adjust themselves to the slow tunnelling motion of the particle
and hence can be treated adiabatically. These modes give rise
to a renormalization of the tunnelling amplitude,
\begin{equation}
\label{eq:nuRenorm}
\tilde\upsilon=\bar\upsilon\,e^{-\sum_l \alpha_l^2/2}.
\end{equation}

\subsection{Non-adiabatic dynamics}
\label{subsec:non_adiabatic}

In contrast to the usual dissipative case, in what follows we
focus on a bath with {\em discrete} low-energy spectrum
$\omega_k=k\delta\omega$, where the level spacing
$\delta\omega$ is fixed.
The adiabatic renormalization of the tunnelling amplitude by
high-frequency modes in \eqref{eq:nuRenorm} does not depend on
the type of the bosonic bath: it is valid for a single oscillator,
a discrete set of oscillators, or a continuum dense
distribution \cite{Spohn:1985}. On the other hand, the low frequency
modes that are smaller or comparable to the tunnelling amplitude
are responsible for a non-adiabatic dynamics of the particle.
Depending on the density of the low-frequency modes, the dynamics
can be quasiperiodic for a few discrete modes or dissipative for a
dense continuum of modes.

\begin{figure}[bt]
\begin{center}
\includegraphics[scale=0.35,angle=0.]{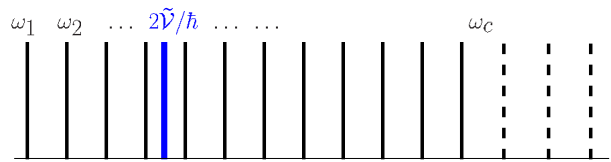}
\end{center}
\caption{(colors online) Bath energy spectrum in the non-adiabatic regime where
several discrete modes have frequencies smaller or comparable to
the tunnelling frequency $2\tilde\upsilon/\hbar$.}
\label{fig:antiadiabatic}
\end{figure}

Let us first separate bath eigenmodes into the low-energy
($\omega_k<\omega_c$) and the high-energy ($\omega_k>\omega_c$) ones.
The high-energy modes renormalize the bare tunnelling amplitude
according to \eqref{eq:nuRenorm}, while the low-energy modes
determine the details of the particle dynamics. The choice of the
cutoff frequency $\omega_c$ is nonessential provided it is much
larger than the frequency of particle tunnelling,
$\omega_c\gg 2\tilde\upsilon/\hbar$ (see figure~\ref{fig:antiadiabatic} and
Appendices A and B).
In this case, the system is described by the Hamiltonian
\eqref{eq:spin-boson} with $\bar\upsilon$ replaced by $\tilde\upsilon$
and $N$ replaced by $N_c$, where $\omega_k$
($k=1,\ldots,N_c$) are the low-energy modes.

%
%
%
%

Next, we apply a polaron unitary transformation
$\hat H'=e^{\hat\sigma_z \hat D} \hat H e^{-\hat\sigma_z \hat D}$
with
$\hat{D} = \sum_{k=1}^{N_c} \alpha_k
\left( \hat a_k^{\phantom{g}}  - \hat a_k^\dagger \right)$,
in which the oscillators are displaced depending on the state of a
particle. The transformed Hamiltonian reads
\begin{equation}\label{eq:spin-boson_LF}
\hat H' =
\tilde\upsilon \,
\left(
    \hat\sigma_- e^{-\hat D}
    +
    \hat\sigma_+ e^{\hat D}
\right)
+
\sum_{k=1}^{N_c} \hbar\omega_k^{\phantom{g}} \hat a_k^\dagger \hat a_k^{\phantom{g}},
\end{equation}
where $\hat\sigma_- = \ket{L}\bra{R}$,
$\hat\sigma_+ = \ket{R}\bra{L}$, and we omitted an unimportant
additive constant in $\hat H'$. For zero coupling $\hat{D}=0$
the tunneling of the free particle is recovered
($\hat\sigma_- + \hat\sigma_+ = \hat\sigma_x$).
The time evolution of $\hat\sigma_\pm$ with respect to $H'$
is given by
\begin{equation}\label{eq:sigma_pm_t}
\hat\sigma_\pm(t) =
\hat\sigma_\pm(0) \pm
\frac{i 2 \tilde\upsilon}{\hbar}
\int_0^t dt'  \, e^{\mp\hat D(t')} \hat\sigma_z(t').
\end{equation}
Substituting $\hat\sigma_\pm(t)$ in the equation of motion for
$\sigma(t) \equiv \langle\hat\sigma_z(t)\rangle$, we obtain
\begin{equation}\label{eq:main_equation}
\frac{d\sigma(t)}{dt}
=
\frac{i\tilde\upsilon}{\hbar}
\left\langle
    e^{\hat D(t)} \hat\sigma_+(t)  - \hat\sigma_-(t)  e^{-\hat D(t)}
\right\rangle
=
- \frac{4\tilde\upsilon^2}{\hbar^2}
\int_0^t dt' \,  \KK(t-t')\, \sigma(t'),
\end{equation}
where
\begin{align}
\label{eq:KK_t}
\KK(t-t') \equiv
\text{Re}\,
\left\langle e^{\hat D(t)} e^{-\hat D(t')} \right\rangle
= \text{Re}\, \left( e^{J(t-t')} \right)
\end{align}
with $J(t) =- \sum_{k=1}^{N_c} \alpha_k^2 (1-e^{-i\omega_k t})$.
Here we have used the initial condition
$\langle \hat\sigma_\pm(0) \rangle = 0$
and the noninteracting blip approximation (NIBA)%
\cite{Leggett:1987,Weiss:2012,Aslangul:1986,Dekker:1987,Porras:2008}
to factorize the average of a product of particle and bath
operators.
The approximation is based on the assumption that the dynamics of
the bath is weakly perturbed by the particle ($\alpha_k^2 \ll 1$),
whereas the back-action of the bath on the particle is taken
into account ($\sum_k \alpha_k^2 \sim N \alpha_k^2$).
%
%
%
%
%
\begin{figure}[t]
\begin{center}
\includegraphics[scale=0.76]{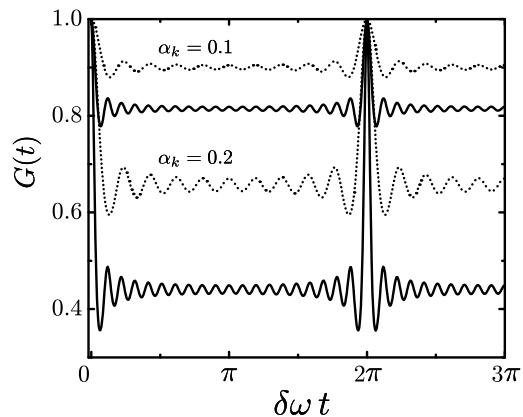}
\end{center}
\caption{The kernel $\KK(t)$ for the bath with $N_c=10$ modes (dotted)
and $N_c=20$ modes (solid curve) and the coupling strength
$\alpha_k=0.1$ (top) and $\alpha_k=0.2$ (bottom). The frequencies of
the modes are assumed equidistant, $\omega_k=k\delta\omega$.}
\label{fig:kernel}
\end{figure}

For a state $\ket{\psi(t)}=c_L(t)\ket{L} + c_R(t) \ket{R}$, the
quantity $\sigma(t)=|c_L(t)|^2-|c_R(t)|^2$ measures the degree of
superposition of $\ket{L}$ and $\ket{R}$ states.
Equation \eqref{eq:main_equation} describes the particle
dynamics in a closed form for a given kernel $\KK(t-t')$
characterizing the bath.
The kernel $\KK(t)$ is shown in
figure~\ref{fig:kernel} for equidistant bath frequencies
$\omega_k = k\delta\omega$ and different number of modes $N_c$
and the coupling strengths $\alpha_k$.
At a given coupling constant $\alpha_k$, for $N_c\sim 1$, the kernel
$\KK(t)$ exhibits oscillations with a small amplitude and period
$\tau_r=2\pi/\delta\omega$ which corresponds to the revival time.
When the number of modes $N_c$ is increased, the kernel $\KK(t)$ decays
at short times with a time constant $(\sum_k \alpha_k^2
\omega_k^2/2)^{-1/2}$ which corresponds to the typical duration of
the revivals occurring after a period $\tau_r$.
To complete the analysis, we note that $\KK(t)$ has also another time
scale $\tau_s$ for high cut-off $N_c$, associated with the fast
oscillations inside the duration of one revival, with frequency
$\sim \sum_k \alpha_k^2 \omega_k$.

%
%
%
%
\begin{figure}[t]
%
\begin{center}
\includegraphics[scale=0.75]{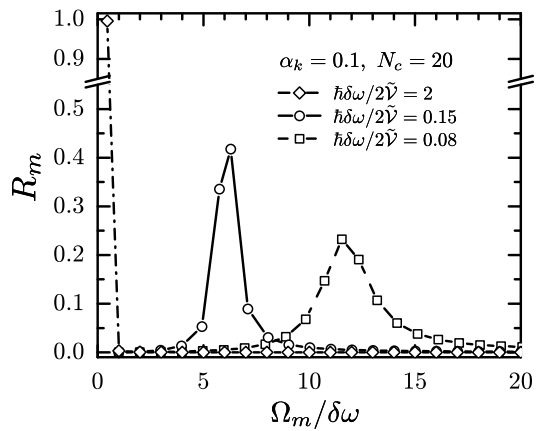}
\end{center}
\caption{Frequency spectrum for a particle coupled to a bath
of $N_c=20$ modes with $\alpha_k=0.1$ and the level spacing
$\hbar\delta\omega/2\tilde\upsilon = 2$ (diamonds), $0.15$ (circles),
and $0.08$ (squares).}
\label{fig:plRmOm}
\end{figure}

In what follows we solve \eqref{eq:main_equation} assuming
equidistant low-energy spectrum of the bath
$\omega_k=k\delta\omega$ ($k=1,\ldots,N_c$).
Taking the Laplace transform of \eqref{eq:main_equation}
we obtain
\begin{equation}\label{eq:sigma-s}
\sigma(s) = \frac{\sigma_0}{s+(4\tilde\upsilon^2/\hbar^2) \KK(s)}
\end{equation}
where $\sigma_0\equiv\sigma(t=0)$ and
$\KK(s) = \sum_{m=0}^\infty c_m s/(s^2+\omega_m^2)$.  The coefficients
$c_m$ are given by
\begin{equation}\label{eq:C_m}
c_m =
e^{-\sum_k  \alpha_k^2}
\sideset{}{'} \sum_{\{ k\}}
\frac{ \alpha_1^{2k_1} \alpha_2^{2k_2} \cdots
\alpha_{N_c}^{2k_{N_c}}  }
{k_1!k_2!\cdots k_{N_c}!}
\end{equation}
where $'$ denotes summation over $k_n\ge 0$ with constraint
$\sum_{n=1}^{N_c} n k_n = m$.
The constraint takes into account the degeneracy of the energy
eigenstate $\hbar\omega_m$ of the bath. Coefficients
$c_m$ obey the sum rule $\sum_{m=0}^\infty c_m = 1$.
%
%
We recall that the coupling of the particle to the bath is assumed to
be small, $\alpha_k^2 \ll 1$, but may vary as a function of $k$ for
different modes $\omega_k$ of the bath.

Equation \eqref{eq:sigma-s} has poles at $s=\pm i\Omega_m$,
where $\omega_m<\Omega_m<\omega_{m+1}$ ($m=0,1,\ldots$). Taking the
inverse Laplace transform of \eqref{eq:sigma-s} we obtain
\begin{equation}
\label{eq:sigma_t} \sigma(t) =
\sigma_0 \sum_{m=0}^\infty R_m \cos(\Omega_m t)
\end{equation}
with
$R_m = \prod_{n=1}^\infty (\omega_n^2-\Omega_m^2) /
\prod'\!{}_{n=0}^{\infty} (\Omega_n^2-\Omega_m^2) $.
Here, $'$ denotes that the term with $n=m$ is omitted in the
denominator of $R_m$.

%
%
%
%
\begin{figure}[t]
\begin{center}
\includegraphics[scale=0.86]{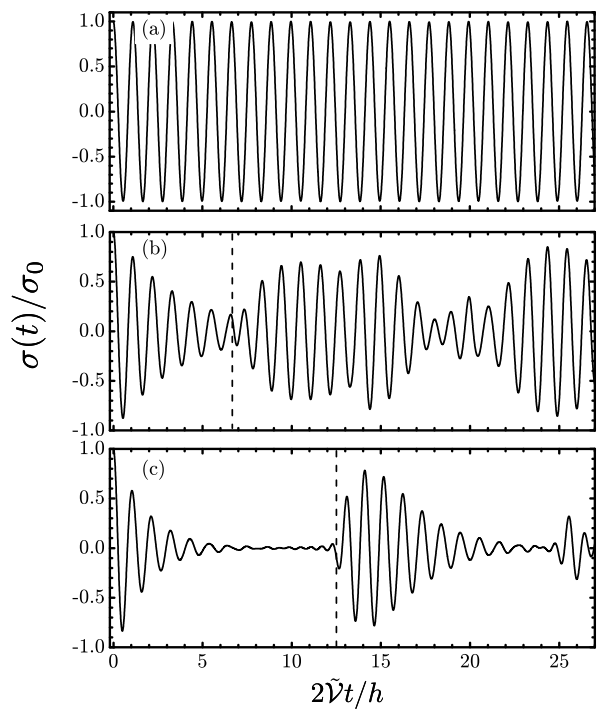}
\end{center}
\caption{Average position $\sigma(t)$ for a particle 
coupled to a discrete bath with $N_c=20$, $\alpha_k=0.1$,
and the level spacing (a) $\hbar\delta\omega/2\tilde\upsilon=2$,
(b) $\hbar\delta\omega/2\tilde\upsilon=0.15$, and (c)
$\hbar\delta\omega/2\tilde\upsilon=0.08$. The corresponding frequency
spectra are shown in figure~\ref{fig:plRmOm}. Dashed lines in (b)
and (c) indicate the onset of revivals at $t = 2\pi/\delta\omega$.}
\label{fig:plSofT}
\end{figure}

A crossover from adiabatic to non-adiabatic dynamics is shown in
figures~\ref{fig:plRmOm} and \ref{fig:plSofT} for a particle
coupled to a bath with $N_c=20$ modes, $\alpha_k=0.1$, and
the level spacing $\hbar\delta\omega/2\tilde\upsilon=2$, $0.15$, and
$0.08$, respectively. The average position of a particle $\sigma(t)$
is shown in figure~\ref{fig:plSofT}.
In the adiabatic case $\hbar\delta\omega/2\tilde\upsilon=2$,
we observe in figure~\ref{fig:plRmOm} that only the lowest frequency
is relevant. It is approximately equal to the renormalized
frequency given by \eqref{eq:nuRenorm} with the sum including
all the modes (see \ref{app:1}).
In this case the dynamics corresponds simply
to coherent oscillations shown in figure~\ref{fig:plSofT}(a).
As the density of the modes is increased, several frequencies
$\Omega_m$ start to contribute, with amplitudes $R_m$ shown in
figure~\ref{fig:plRmOm}.
In the weak coupling regime which we consider, the particle still
oscillates between the two minima with the frequency
$2\tilde\upsilon/\hbar$ corresponding to the fast oscillations in
figures~\ref{fig:plSofT}(b) and (c).
The amplitude of these oscillations initially decays as the bath
modes are populated and the energy is transferred from the
particle to the bath.
The decay time is
$\tau_d \propto \hbar^2/\tilde\upsilon^2 \sqrt{(\sum_k
\alpha_k^2 \omega_k^2)/(\sum_k \alpha_k^2)}$. However, after time
$\tau_r = 2\pi/\delta\omega$, the populated bath modes start to
feed energy back to the particle and revivals of oscillations take
place.
From that point on, we have two different behaviors depending on
the ratio $\tau_d/\tau_r$.
For $\tau_d \lesssim \tau_r$, the dynamics of a particle has a
form of a quasiperiodic beating instead of a decay.
Reducing $\tau_d \ll \tau_r$,  the dynamics exhibits again
a decay after a revival of the oscillation amplitude.
For a dense continuum of bath modes
($N_c\to\infty$, $\delta\omega\to 0$) the revival time is infinite,
$\tau_r\to\infty$. In this case the bath cannot feed significant
amounts of energy back to the particle and one recovers
exponentially damped oscillations characteristic for Ohmic
dissipation.

\section{Josephson junction ring with a weak element}
\label{sec:1DJJ}

Here we show how the dynamics of the quantum tunnelling between the
two counterpropagating supercurrent states discussed in
section~\ref{sec:QPS} can be mapped to the spin-boson model of
section~\ref{sec:model} when the electric modes of the ring
are taken into account.

First we cast \eqref{eq:1DJJ_1} and \eqref{eq:1DJJ_2} in the form in which the
coupling of $\theta$ to electric modes of the ring is manifest \cite{Rastelli:2013}.
We take as independent variables the phases between  Josephson junctions,
$\varphi_n \equiv \phi_n$ for $n=1,\dots,N-1$, the average phase
$\varphi_0 \equiv (\phi_0+\phi_N)/2$ and the phase difference $\theta$
across the weak element.
Since $\varphi_n$ is periodic on the effective lattice $n=0,\dots,N-1$ composed of $N$
elements, it can be Fourier transformed as
$\varphi_n =(1/\sqrt{N}) \sum^{N-1}_{k=0} \varphi_k \exp(i 2\pi n k / N)$
where $\varphi_{N-k} = \varphi_k^*$.
The real and imaginary parts $\varphi'_k$ and $\varphi''_k$ of
$\varphi_k$ give rise to even and odd modes, respectively.
After the substitution in \eqref{eq:1DJJ_1} and \eqref{eq:1DJJ_2}, we find that only
$\varphi''_k$ couple to $\theta$ while $\varphi'_k$ describe a
set of decoupled harmonic oscillators. 
Since $\varphi''_{N-k} = - \varphi''_k$, only half of the modes
are independent; we label these modes with $k$, $1 \leq k \leq k_{\rm max}$,
where $k_{\rm max}=\lfloor (N-1)/2 \rfloor$ and $\lfloor x \rfloor$
is the integer part of $x$.
The Euclidean Lagrangian in the imaginary time reads
$\Lg = \Lg_0 + \Lg_{\rm int}$ where
\begin{equation}\label{eq:L_theta}
\Lg_0
=
\frac{\hbar^2\dot\theta^2}{16 E_{\tilde C}}
- \bar E_J \cos(\theta+\delta_B) +
\frac{E_L}{2} (\theta - N \delta_B)^2
\end{equation}
and
\begin{equation}\label{eq:delta_L}
\Lg_{\rm int}= \sum_{k=1}^{k_{\rm max}}
\left\{
    \frac{\mu_k}{2}
   \dot{X}_k^2
    +
    \frac{\mu_k \omega_k^2}{2}
    \left[ X_k - \left(\frac{\omega_p^2}{\omega_k^2} - 1 \right) \frac{f_k}{\mu_k} \theta \right]^2
\right\},
\end{equation}
with $\delta_B = 2\pi(\Phi_B/\Phi_0)/(N+1)$ and $X_k = \varphi''_k - (f_k/\mu_k) \theta$.
Here, $\mu_k=(8E_J/\omega_k^2)\sin^2(\pi k/N)$,
$f_k =(2E_J/\sqrt{N}\omega_p^2)\sin(2\pi k/N)$, and $\omega_p=1/\sqrt{L_J C}$.

The Lagrangian $\Lg_0$ describes the phase $\theta$ in a double-well
potential with two degenerate minima at half flux quantum
($L\gg \bar L_J$, where $L=N L_J$ is the effective inductance
of the ring; $\bar L_J$ is inductance of the weak junction).
The minima correspond
to the counterpropagating supercurrent states that enter the
spin-boson model and which are coupled by the quantum phase tunnelling,
cf. section~\ref{sec:model} and figure~\ref{FIG_1dw_oscillators}
\cite{Mooij:1999,MLG:2002,Rastelli:2013}.
Josephson junctions
in the chain give rise to a renormalization of the charging energy
of the weak element $E_{\tilde C}=e^2/2\tilde C$, where
\begin{equation}\label{eq:C_ren}
\tilde C
=
\bar C + \frac{C}{N} +
\frac{C_0}{2}
\left(
    1 + \frac{1}{N} \sum_{k=1}^{k_{\rm max}}
    \frac{\cos^2(\pi k/N)}{\sin^2(\pi k/N)+C_0/4C}
\right).
\end{equation}
%
By taking the thermodynamic limit $N \rightarrow \infty$
in (\ref{eq:C_ren}), we recover the renormalization of the
capacitance of the weak junction as obtained in
\cite{Korshunov:1989} (see further discussions in \cite{Rastelli:2013}).
On the other hand, when the capacitance to the ground is small,
$N\sqrt{C_0/C} \ll 1$, we obtain
$\tilde{C} = \bar{C} + C/N  +C_0/2 + C_0 (N-1)(N-2)/12N$
which is in agreement with \cite{Ferguson:2013}.
This result can be obtained by setting $C_0=0$
in the sum over $k$ in \eqref{eq:C_ren}.

The Lagrangian $\Lg_{\rm int}$ in \eqref{eq:delta_L} contains the harmonic modes in the ring whose
dispersion relation  reads\cite{Rastelli:2013}
%
\begin{equation}\label{eq:wk}
\omega_k
=\frac{\omega_p \sin(\pi k/N)}{\sqrt{\sin^2(\pi k/N) + C_0/4C}}.
\end{equation}
%
Note that the potential term in \eqref{eq:delta_L} does not
confine $\theta$ because it depends on the relative coordinates
with respect to the bath degrees of freedom.
Moreover, we note that the ground capacitance plays a crucial role:
For $C_0 = 0$ the dispersion relation becomes flat with $\omega_k = \omega_p$
and the weak junction is decoupled from the electric modes of the ring,
see \eqref{eq:delta_L}.
%
In that case, the only effect of the Josephson ring is the presence of the adiabatic
confining potential in $\Lg_0$ associated with the ring's inductance.
This result is in agreement with previous works \cite{MLG:2002,Rastelli:2013,Ferguson:2013}
in which it was shown that the modes of the Josephson chains are decoupled
from the weak element in the harmonic approximation and for $C_0=0$.


The harmonic modes of the ring can be integrated out using the
Feynman-Vernon influence functional in the real-time path integral
approach.
For the model  \eqref{eq:spin-boson}, the resulting influence action which governs the dynamics of
the two levels is a functional of the spectral density of the modes
\begin{equation}\label{eq:J_omega}
F(\omega)
=
\frac{\hbar}{\pi} \sum_k \alpha_k^2 \omega_k^2
\delta(\omega - \omega_k).
\end{equation}
In a similar way, the linear coupling of the phase difference $\theta$ at the weak
element to an ensemble of harmonic oscillators affects the
dynamics of $\theta$ only through $F(\omega)$, regardless of the
details of the bath \cite{Weiss:2012,Leggett:1987}.
Hence, from the knowledge of the coupling constants $\alpha_k$ and
the spectrum $\omega_k$ one can analyze the real-time dynamics of
the quantum tunnelling between the two low-energy states in a
double-well potential of equation~\eqref{eq:L_theta} using the effective
spin-boson model as described in section~\ref{sec:model}.

Instead of carrying out a calculation in the real-time formalism,
we can proceed with the imaginary-time one and make use of a
relation\cite{Weiss:2012}
\begin{equation}\label{eq:K_omega}
K_l
=
\frac{2}{\pi}
\int^\infty_0
\!\!\!\!
d\omega
\frac{\nu_l^2 F(\omega)}{\omega(\nu_l^2+\omega^2)}
=
\frac{2\hbar}{\pi^2}
\sum_k \alpha_k^2
\frac{\nu_l^2 \omega_k}{\nu_l^2+\omega_k^2}
\end{equation}
between $F(\omega)$ and the kernel $K_l=K(\nu_l)$ of the
imaginary-time effective action ($\nu_l = 2\pi l/\beta$
are the Matsubara frequencies).
Kernel $K_l$ can be obtained from the partition function of the
system which is given by imaginary-time path integral over closed
trajectories $\theta(0)=\theta(\beta)$ and
$X_k(0)=X_k(\beta)$:
\begin{equation}
\ZZ_{\rm tot}= \oint \DD\theta\DD X\, e^{-(S_0+S_{\rm int})/\hbar},
\end{equation}
where $S_0[\theta]=\int_0^{\beta}d\tau \Lg_0[\theta]$ and
$S_{\rm int}[\theta,X]=\int_0^{\beta}d\tau\Lg_{\rm int}[\theta,X]$.
After integrating out bath degrees of freedom, one obtains
$\ZZ_{\rm tot}=\ZZ_h\times \ZZ$, where
$\ZZ_h=\prod_k [2\sinh(\beta\omega_k/2)]^{-1}$
is the partition function of harmonic oscillators and
\begin{equation}
\ZZ = \oint \DD\theta\, e^{-(S_0+S_{\rm inf})/\hbar}
\end{equation}
is the partition function of the particle interacting with
the bath. The interaction is included in the influence action
\begin{equation}\label{eq:K}
S_{\rm inf}[\theta]
=
\frac{1}{2}\int_0^{\beta}  \!\!\!\!\!
d\tau d\tau' \theta(\tau)K(\tau-\tau')\theta(\tau')
=
\frac{1}{\beta} \sum_{l=1}^\infty K_l |\theta_l|^2,
\end{equation}
where $\theta_l = \int_0^{\beta}d\tau\, \theta(\tau)
e^{i\nu_l\tau}$.
After integration of the harmonic modes in $\Lg_{\rm int}$,
we obtain
\begin{equation}\label{eq:delta_S}
S_{\rm inf}[\theta]
=
\frac{1}{\beta} \sum_{l=1}^\infty |\theta_l|^2
\sum_{k=1}^{k_{\rm max}}
\frac{\nu_l^2}{\nu_l^2+\omega_k^2}
\left(
\frac{f_k^2 \omega_p^4}{\mu_k \omega_k^2}
\right)
\left( 1 - \frac{\omega_k^2}{\omega_p^2} \right)^2,
\end{equation}
and using \eqref{eq:K_omega} and \eqref{eq:K} we extract the
coupling constants:
\begin{equation}\label{eq:alpha_k}
\alpha_k
=
\frac{\pi}{\sqrt{N}}
\left( \frac{E_J}{\hbar\omega_k} \right)^{\! 1/2}
\!\!
\left( 1 - \frac{\omega_k^2}{\omega_p^2} \right)
\cos(\pi k/N).
\end{equation}

Equations \eqref{eq:wk} and \eqref{eq:alpha_k} for
the frequencies $\omega_k$ of electric modes in the loop and the
coupling constants $\alpha_k$, respectively, complete the mapping
of the Josephson junction ring with a weak element to a
generic spin-boson model of section~\ref{sec:model}.


For a non-zero capacitance to the ground and large number of junctions
in the ring, $N\sqrt{C_0/C} \gg 1$,
the dispersion at low frequencies is linear,
$\omega_k \approx (2\pi k/N)\omega_0$
($\omega_0 = 1/\sqrt{L_JC_0}$).
In this case, the coupling constants at low frequencies are given by
\begin{equation}\label{eq:alpha_k_2}
\alpha_k = \frac{1}{2}\sqrt{\frac{R_q}{Z_0}}\frac{1}{\sqrt{k}}
\qquad
(k<N_c),
\end{equation}
where $R_q=h/4e^2$ is the quantum resistance and
$Z_0=\sqrt{L_J/C_0}$ is the low-frequency
transmission-line impedance of the ring.
The cutoff frequency $\omega_c=\omega_{k=N_c}$ with
$\sin(\pi N_c/N)=\sqrt{C_0/4C}$ discriminates between a linear
(low-frequency) and a nonlinear (high-frequency) part of the
spectrum.
As long as  $2\tilde{\upsilon} < \hbar\omega_p$,
this frequency also divides the low-frequency modes 
responsible for the details of the phase dynamics from the
high-frequency modes which only renormalize the phase slip
amplitude.

Before we conclude this section, let us also consider the case
of the small capacitance to the ground, $N\sqrt{C_0/C} \ll 1$.
The coupling constants in this case are given by
$\alpha_k \approx (1/4\pi) (C_0/C) (E_J/8E_C)^{1/4} N^{3/2}/k^2$.
The effective Hamiltonian of the weak junction coupled to
the electric modes of the ring is given by
$\hat H_{eff} = -4E_{\tilde C} \partial_{\hat\theta}^2 + V(\hat\theta)
+ (\hat\theta/\pi) \sum_k \alpha_k \hbar\omega_k (\hat a_k^\dagger + \hat a_k)
+ \sum_k \hbar\omega_k \hat a_k^\dagger \hat a_k.$
By applying the unitary transformation $\hat U^\dagger \hat H_{eff} \hat U$
where $\hat U = \exp[(\hat\theta/\pi)\sum_k \alpha_k(\hat a_k - \hat a_k^\dagger)]$
we can cast the Hamiltonian in the form in which the coupling is expressed
in terms of momenta rather than coordinates. We obtain
$\hat U^\dagger \hat H_{eff} \hat U =
-4E_{\tilde C} \partial_{\hat\theta}^2 + V(\hat\theta)
-\sum_k \lambda_k (\hat a_k - \hat a_k^\dagger)
\partial_{\hat\theta}
+ \sum_k \hbar\omega_k \hat a_k^\dagger \hat a_k$,
with $\lambda_k = (8/\pi^2) E_{\tilde C} (C_0/C) (E_J/8E_C)^{1/4} N^{3/2}/(2k)^2$.
The obtained coupling term $-\sum_k \lambda_k (\hat a_k - \hat a_k^\dagger)
\partial_{\hat\theta}$ between the weak element and the modes of the ring is
in agreement with the results of Ferguson {\it et al.} \cite{Ferguson:2013}.

\subsection{Discussion of the experimental observability}

In the following we analyze the feasibility of achieving a
non-adiabatic dynamics in realistic superconducting rings
made of Josephson junctions.
Since the capacitance of the junction is proportional to the
cross section area while inductance is inversely proportional
to it, we have $L_J C = \bar L_J \bar C$.
In this case, the condition $NL_J \gg \bar L_J$ for the system
to be in a well-defined flux state implies $\bar C \gg C/N$
and the renormalization of the capacitance of the weak element
in \eqref{eq:C_ren} is negligible.
The hierarchy of energy scales
$N \upsilon \ll \tilde\upsilon \ll E_L$  discussed in
section~\ref{sec:QPS} gives the upper limit of the ring length,
$N \ll N_a, N_b$, where
$N_a=\sqrt{L_J/\bar L_J}\exp[(4/\pi)R_q(Z_J^{-1}-\bar Z_J^{-1})]$
and
$N_b \approx 3.5 (\bar L_J/L_J)\sqrt{R_q/\bar Z_J}
\exp[(4/\pi)R_q/\bar Z_J]$ with $Z_J=\sqrt{L_J/C}$ and
$\bar Z_J=\sqrt{\bar L_J/\bar C}$.
These conditions are not very restrictive and can be met in
realistic devices, as demonstrated experimentally in the
fluxonium superconducting chain with $N=43$ Josephson
junctions in series \cite{Manucharyan:2009}.
In addition to the previous conditions, for non-adiabatic
phase dynamics to occur the lowest frequency of electric modes
has to be smaller than the qubit level splitting,
$\hbar\delta\omega= 2\pi\hbar\omega_0/N < 2\tilde\upsilon$.
This can be achieved, e.g., by making the ground capacitance
larger than a certain threshold,
$C_0 > (\pi\hbar/N\tilde\upsilon)^2 L_J^{-1}$.

%
%
%
%
\begin{figure}[t]
\begin{center}
\begin{tabular}{p{0.5\textwidth}p{0.5\textwidth}}
  \vspace{0pt}\hspace*{8mm}\includegraphics[scale=0.7]{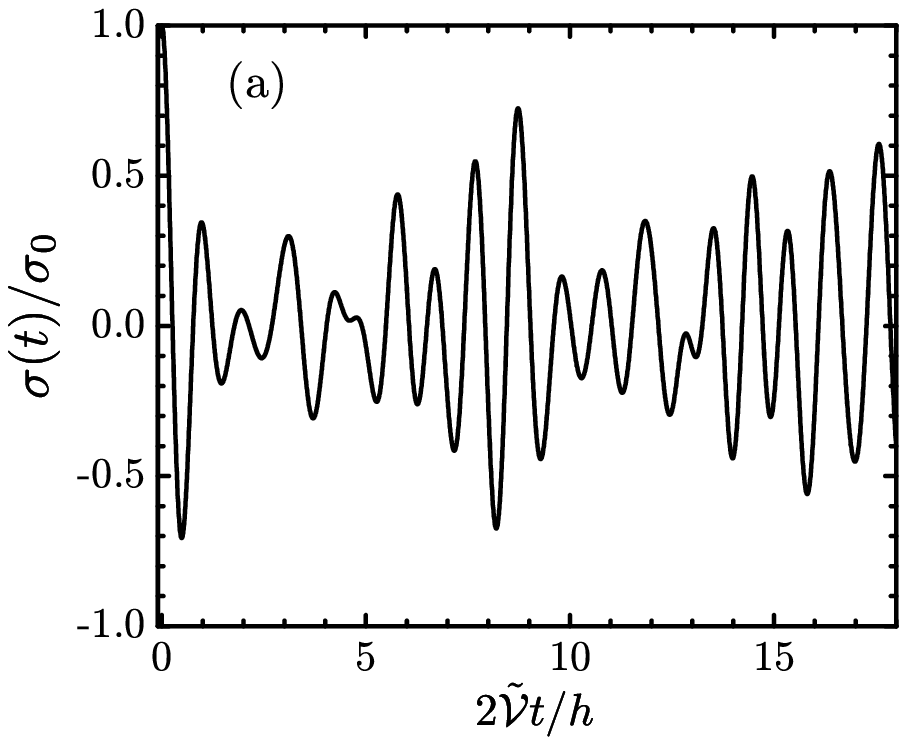} &
  \vspace{0pt}\hspace*{0mm}\includegraphics[scale=0.7]{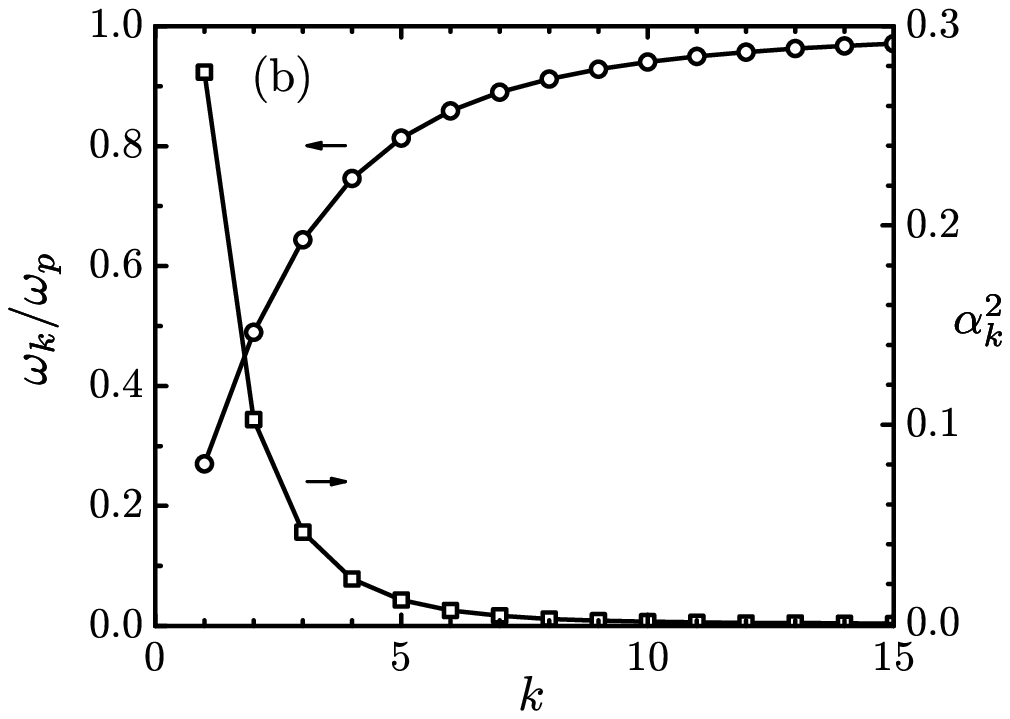}
\end{tabular}
\end{center}
\caption{(a) Non-adiabatic dynamics of a flux qubit made
of a Josephson junction chain with a weak element.
Parameters are $\bar C/C=0.1$, $C_0/C=0.05$, $Z_J/R_q=0.18$,
$N=100$, and $\hbar\omega_p/\tilde\upsilon=3$.
(b) Dispersion $\omega_k$ (circles, left axis)
and the coupling constants $\alpha_k^2$ (squares, right axis)
of the modes in the chain.}
\label{fig:plSofTjjChain}
\end{figure}

As an example, we take $N=100$ junctions in series,
$\bar C/C=0.1$, $C_0/C=0.05$, and $Z_J/R_q=0.18$.
The dispersion of the modes and the coupling constants are given by
\eqref{eq:wk} and \eqref{eq:alpha_k}, respectively,
see figure~\ref{fig:plSofTjjChain}(b).
The non-adiabatic dynamics of the qubit is shown in
figure~\ref{fig:plSofTjjChain}(a) obtained by numerical solution of
\eqref{eq:main_equation}. At half flux quantum, the neighboring
phase-slip states carry the counterpropagating persistent currents
of the same magnitude and $\sigma(t)$ is proportional
to the average current through the loop, $\sigma(t)\propto I(t)$.
The dynamics exhibits the same qualitative
features (initial decay and revivals) as discussed
in section~\ref{sec:model} for a generic model with equidistant
spectrum of the modes and constant coupling
of the phase to the bath degrees of freedom. When the number
of junctions or the strength of the coupling $\alpha_k$ is increased (e.g.,
by increasing the capacitance $C_0$ to the ground), the number
of electric modes that are coupled to the phase increases and
the transition to coherent non-adiabatic dynamics takes place.
%
Recent experiments reported the fabrication
of long Josephson junction chains comparable in  order of
magnitude to our example and operating as linear
``superinductance''  elements in which quantum phase slips are
suppressed \cite{Masluk:2012}.
Dispersion of the modes in this system has also been
measured. In addition, Josephson junction
chains in the ladder geometry have been studied
experimentally \cite{bell_quantum_2012}. In this system, quantum
phase tunnelling is prevented at the topological level
which opens the route towards realization of long Josephson
junction chains behaving as perfect inductances.

The condition for the observation of the
quasiperiodic dynamics is that the relaxation and dephasing times of
a qubit are larger than the revival period $\tau_r$.
Estimating $\tau_r \sim 2\pi/\delta\omega$ with
$\delta\omega=0.01\omega_p$ and $\omega_p \sim 1$GHz, we obtain
$\tau_r$ of order of $\mu$s. This is well below the measured relaxation
and dephasing times which can approach hundreds and tens of $\mu$s,
respectively, in the present devices \cite{Pop:2014,Vool:2014}.
Moreover, the experimental resolution for monitoring the qubit which
has been achieved so far can be around of hundred of nanoseconds
pointing out that the observation of the quasiperiodic dynamics is
within the reach of the present technology.

\section{Conclusion}
\label{sec:Summary}

In conclusion, we have studied quantum dynamics between two macroscopic supercurrent
states in superconducting one-dimensional rings with a weak element and threaded by a
magnetic flux.
For sufficiently large system size, we have found that the quantum dynamics can be more
complex than the usual coherent oscillations between the two states characterized by the
quantum phase tunnelling amplitude.
Such a dynamics emerges due to the coupling between the phase difference at the
weak element and the intrinsic electrodynamic modes in the
homogeneous part of the ring.
We have obtained the spectrum of the modes in the ring and the corresponding coupling
constants and have shown that in the non-adiabatic regime the dynamics of the system is
quasiperiodic with exponential decay of oscillations at short times followed by
oscillation revivals at later times.
Revivals can be observed in a typical flux qubit setup in which the state of the qubit is
measured.
We have discussed the experimental feasibility to observe the quasiperiodic dynamics and
revivals in realistic systems with large number of Josephson junctions in series or
in systems with a finite charging energy of the islands between the junctions
due to a non-zero capacitance to the ground.

Recent experiments have shown that a larger number of degrees of freedom is not
necessarily penalized by decoherence
\cite{Manucharyan:2012,Nigg:2012,Ferguson:2013,Peropadre:2013,Vool:2014,Pop:2014},
thus opening the possibility to explore novel dynamic regimes beyond the two-level's one.
Observation of a quasiperiodic dynamics would be important for understanding the
mechanisms of decoherence in large quantum circuits as well as intrinsic limits on
coherence posed by the circuit itself.
Our results can also be of interest for the design of models with a tunable fictitious
dissipation or, for instance, to achieve controlled quantum evolution in superconducting
qubits by engineering the parameters of the Josephson junction circuits.
This motivates future studies of flux qubits realized in large superconducting circuits
with a more complex topological structure \cite{bell_quantum_2012}.
The approach we use is not restricted to superconducting circuits and can be readily
generalized for other situations in which the intrinsic bosonic degrees of freedom couple
to the phase, like in quasi-1D superfluid condensates \cite{Wright:2013,Amico:2013,Weiss:2015}.

\ack

We acknowledge Jon Fenton for valuable comments and fruitful
discussions. The research was supported by the EU FP7
Marie Curie Zukunftskolleg Incoming Fellowship Programme,
University of Konstanz (grant No. 291784).
WB and GR acknowledge financial support by the DFG
through grant No. BE 3803/5.
MV acknowledges support by the Serbian Ministry of Science, project No. 171027.

\appendix

\section{Phase dynamics and the adiabatic regime}
\label{app:1}

Here we analyze the relation between the adiabatic renormalization
of the amplitude in \eqref{eq:nuRenorm} and the time dynamics
of the phase given by \eqref{eq:main_equation}.
Let us start with the bare Hamiltonian in \eqref{eq:spin-boson}
in the regime in which all the frequencies satisfy the adiabatic
condition $\hbar \omega_1 = \hbar\delta\omega \gg 2\bar\upsilon$.
Then, by applying the same steps of section~\ref{subsec:non_adiabatic},
we obtain \eqref{eq:sigma-s} with $\bar\upsilon$ replacing
$\tilde\upsilon$ and for $N$ harmonic oscillators.
At low $s \ll \delta\omega <  \omega_m$  (long time intervals), we
can approximate $G(s)$ in the denominator of \eqref{eq:sigma-s}
by its first term:
\begin{equation}\label{eq:sigma_laplace_ren}
\frac{\sigma(s)}{\sigma_0}
\approx
\frac{1}
{ s + (4\bar\upsilon^2/\hbar^2) c_0 /s }
=
\frac{s}
{s^2 + 4 \tilde\upsilon^2 / \hbar^2 } \, ,
\end{equation}
where $c_0 = \exp(-\sum_{k \geq 1} \alpha_k^2)$.
Thus, there is a single pole $2\tilde\upsilon/\hbar$ at low-frequencies
(see figure~\ref{fig:plRmOm} for $\hbar\delta\omega/2\tilde\upsilon = 2$)
whereas the other poles are relevant only at higher frequencies
$(\sim \delta\omega)$.
In the time domain, \eqref{eq:sigma_laplace_ren} corresponds
to an oscillatory two-levels evolution with a renormalized
frequency $2 \tilde\upsilon/\hbar$ as compared to the bare frequency in
\eqref{eq:spin-boson} and we recover the adiabatic phase
dynamics.

\section{Independence on the cut-off $N_c$}
\label{app:2}

Now we demonstrate that the solution associated to the effective
spin-boson model in \eqref{eq:spin-boson_LF} corresponds to
the low-frequency solution of the bare spin-boson system in
\eqref{eq:spin-boson} and that such a solution is independent
of the high-frequency cut-off $\omega_c$ provided that $\omega_c$
is chosen sufficiently large
$\omega_c \gg \delta\omega \sim \tilde\upsilon$.
This is equivalent to show that the product $\tilde\upsilon^2  G(s)$
in \eqref{eq:main_equation} does not change at
low-frequencies $s \ll \omega_c$.

First, we shift the cut-off $\omega_c'= \omega_c+\delta\omega$,
namely $N_c' = N_c+1$,  so that we have to re-scale all the
parameters  accordingly.
Recalling that
$\tilde\upsilon/\bar\upsilon = \exp(-\sum_{k=N_c+1}^{\infty} \alpha_k^2/2 )$,
we obtain for the renormalized amplitude
\begin{equation}
\tilde\upsilon'  = \tilde\upsilon \exp(\alpha_{N_c+1}^2/2) \, ,
\end{equation}
whereas for the coefficients $c_m$ we have
\begin{equation}\label{eq:new_constraint}
c_m'= e^{-\sum_{k=1}^{N_c+1}  \alpha_k^2}
\, \,
\sideset{}{'} \sum_{\{ k\}}
\frac{ \alpha_1^{2k_1} \alpha_2^{2k_2} \cdots
\alpha_{N_c+1}^{2k_{N_c+1}}  }
{k_1!k_2!\cdots k_{N_c+1}!} \, ,
\end{equation}
with the new constraint $\sum_{n=1}^{N_c+1} n k_n = m$.
Importantly we notice that, for every $m$ such that $m \leq N_c$,
the sum for the two sets of coefficients  $\{ c_m \}$ and
$\{ c_m' \}$ satisfies the same constraint because the term
$n=N_c+1$ is not involved in \eqref{eq:new_constraint}
as it can not satisfy the constraint
$n k_n = N_c$ for any integer $k_n$.
Therefore, we have simply
\begin{equation}
c_m'= \exp \left( -\alpha_{N_c+1}^2 \right)  c_m \quad \text{for}
\quad m \leq N_c.
\end{equation}
In this way  we have demonstrated that the product
\begin{equation}
\tilde\upsilon^2 c_m = \mbox{const.} \quad \mbox{for} \quad m \leq N_c,
\end{equation}
that is, it does not change under the shift of the cut-off.

As second step, we demonstrate that the latter property implies
that the product $\tilde\upsilon^2 G(s)$ is also invariant at low
frequency.
Similarly as in \ref{app:1}, $G(s)$ has a natural
time-scale separation between the (slow) dynamics of the phase and
the (fast) dynamics of the oscillators at high-frequency.
At low frequency $s \ll  \omega_c = N_c \delta\omega$,
we can approximate the product
\begin{equation}\label{eq:G_s_low}
\tilde\upsilon^2 G(s)
=
\sum_{m=0}^{\infty}
\frac{\tilde\upsilon^2 c_m s}{s^2+\omega_m^2}
\approx
\sum_{m=0}^{N_c}
\frac{(\tilde\upsilon^2 c_m) s}{s^2+\omega_m^2},
\end{equation}
since $s \ll \omega_m$  and the coefficients $c_m$ also decrease
for $m>N_c$.
Because the low-frequency form of $G(s)$ involves only  the
coefficients $c_m$ with $m \leq N_c$, the product
$\tilde\upsilon^2 G(s)$ is indeed invariant under a variation of
the frequency cut-off.

%
%
%
%
%

\section*{References}

\bibliographystyle{iopart-num}
\bibliography{references}

\end{document}